# Analytic Model of Variable Characteristic of Coefficient of Restitution and Its Application to Soccer Ball Trajectory Planning

**Hwan-Taek Ryu [1], Byung-Ju Yi [1], and Younghun Kwon [2]**

[1] *Department of Electronic System Engineering, Hanyang University, Ansan, Korea 426-791*

[2] *Department of Applied Physics, Hanyang University, Ansan, Korea 426-791*

In this article, we investigate the behavior of the coefficient of restitution (COR) which is an important parameter in many impact-related fields. In many cases, the COR is considered as a constant value, but it varies according to many variables. In this paper, we introduce an analytical variable COR model considering aero dynamics along with its verification through experiment. To introduce and analyze the variable characteristic of the COR model, the collision phenomenon between a pendulum and two kinds of ball is employed as an example and aerodynamics such as drag force is considered for analyzing the after-effect of the collision. Collision velocity of the pendulum, dynamic parameters of colliding bodies, contact time, drag coefficient, the air density, and the cross-sectional area of the ball are found as the typical variables of analytical COR model. This observation generalizes the result in previous researches. To verify new COR model, the travel distances for the curve-fitted constant COR model and the curve-fitted variable COR model are compared through simulation and experiment. Moreover, comparison between constant COR and variable COR is presented in several point of views. Finally, using the variable COR model, the travel distance of the ball for collision velocity, which is beyond the curve-fitted range, is estimated.

PACS number: 45.50.Tn: Collisions classical mechanics, 46.55.+d :Mechanical contacts

Keywords: Variable coefficient of restitution (COR), Impulse, Aerodynamics.

# I. INTRODUCTION

It is impossible for us to think any ordinary phenomenon or circumstances without impact or collision. Punching, hitting, kicking and so on in sports action, and hammering, grasping, running and so on in daily action are examples of impact. In impact phenomenon, there are some cases that the bigger impact is desirable such as punching, hitting, and hammering. On the contrary, there are some cases that the smaller or adequate impact is desirable such as kicking, grasping, and running.

The impulse is a quantitative measure of impact phenomenon. Therefore, the impulse should be measured or estimated precisely although it is difficult to control. The impulse has been considered profoundly not only in sports area, but also in robotics field. There exist several researches analyzing the external and internal impulse to relate the external load to the joint load [1-4]. There were several papers that not only concentrate on the phenomenon right after the collision but also the entire tendency of the object travel trajectory [5-7].

In usual mechanics, the coefficient of restitution (COR) has been also considered as a constant value when calculating the impact or impulses [8]. Recently, however, there have been many researchers to demonstrate that the COR has variable characteristic with respect to the collision velocity, contact material, and contact area. Burgeson [9] studied variable characteristic of COR depending on the collision velocity for several pure metals based on the fundamental and simple mathematics, and achieved experimental results. The American Society for Testing and Materials (ASTM) [10] suggested a systematic method for testing the COR for collision between two kinds of the balls and two kinds of the plate, and also concluded that COR was largely dependent on the collision velocity. Andersen et al. [11] studied the relationship between the contact areas and COR by using a biomechanical foot model such as toe kick and instep kick, but there was not enough comparison between theoretical estimation and experimental results. Goff et al.[12] analyzed the travel tendency of the one kind of soccer ball considering aerodynamics such as drag coefficient and Magnus effect for large velocity and large travel distance. Wadhwa [13] conducted an analysis for rebound resilience on one dimension free falling environment based on the fundamental COR equation. Then, he experimentally measured the COR data by using audio signal for collision of two kinds of the ball.

Also, many researchers have proposed the variable COR characteristics based on the finite-element analysis. Chiu et al. [14] and Hill [15] introduced the finite-element model to analyze the variable COR and stress characteristics for golf ball and compared with the experimental result by changing plate thickness. Kanda et al. [16] conducted the 3 dimensional finite-element analysis for impact between a bat and a rubber baseball, and then found the effect between the thickness or stiffness of the bat and rubber ball on COR. Price et al. [17] proposed the dynamic mechanical analysis for two thermoformed soccer balls based on finite-element modeling. The analysis was verified by experimental impact test in several items such as COR and contact time. Sissler et al. [18] presented the explicit finite-element tennis ball model considering the rubber core model and fabric cover part. The analyzed model is compared with experiment for COR value, impact time, and deformation values. Ranga et al. [19] introduced the finite-element modeling to replicate the quasi-static and dynamic behavior of a solid ball and the accuracy between the analytical value and the experiment for stress and COR is discussed. Aryaei et al. [20] research the effect of size, material of metal ball, and material of metal sheet on COR based on a finite-element model for elasto-plastic collision in free falling situation. The dynamic analysis of collision is conducted by using ANSYS.

Besides, there existed several papers using different methods. Mills and Nguyen [21] studied the behavior of the impact force between the manipulator and environment during the contact time by using a singular perturbation theory approach. Yigit [22] performed a research on the impact response for 2-DOF manipulator. As an impact modeling, hertzian contact law is introduced and its simulation result for different rigidity of the manipulator links is presented. Gerl and Zippelius [23] suggested that the COR is the function of the relative velocity between environment and the elastic disks based on Hertz approach in the quasistatic limit. Cross presented the dynamic hysteresis curve to show how energy is lost during the collision for several types of the ball and compared the energy loss by using simple measurement of COR [24]. He also presented the research of COR for horizontal axis by changing the position and orientation of the ball by changing the inclined surface [25].

There were other viewpoints that the COR varies on different factors except the factors mentioned above. Walton and Braun [26] studied several factors that could affect variable characteristic of the COR in particle level. They argued that viscosity, temperature, and stress can affect the fixed friction and restitution coefficients. Kawabara and Kono [27] dealt with the relative energy loss caused by the visco-elastic material and derived the general expression of COR as a function of elastic constant and collision velocity in case of the collision between two spheres. First, he constructed dynamic equation of the collision considering several material parameters, and then achieved COR model. As a result, the comparison of COR between experiment and calculation with more than four different material, size, and mass is conducted, and it is deducted that there exist the tendency that (1-e) is proportional to 5 square root of velocity. Brody [28] proposed the dependence of the court speed on the coefficient of friction and coefficient of restitution for tennis ball and conducted an analysis of theoretical relationship between the court speed and speed of the game. Kagan and Atkinson [29] suggested that the COR of baseball also varies according to the humidity of environment by conducting experiment.

Even though there have been many studies on the variable characteristics of COR, there does not exist a clear picture for analytical model of COR having variable characteristics. Thus, differently from other researches, the goal of this research is the suggestion of an analytical model of COR having variable characteristics in the impact. For correcting Choi and Yi [7]'s work based on constant COR model, this paper is focused on the estimation of more accurate travel distance of kicking balls based on variable COR. Two kinds of ball are employed to kicking experiment.

Aerodynamic effect (i.e., drag force) is also taken into account. We develop the analytical COR model as a function of collision velocity of the pendulum, dynamic parameters of colliding bodies, contact time, drag coefficient, the air density, and the cross-sectional area of the ball, which was known as the typical variables of COR model. Our approach is more general as compared to previous researches. For two kinds of ball, validity of our COR model is proved by comparing the travel distance between experiment and the simulation model. The variable COR is compared with constant COR in several viewpoints. Finally, using the curve-fitted variable COR model, the travel distance of the ball for collision velocity, which is beyond the experimental range of the experimental equipment, can be estimated. This analysis and experimental results for analytical COR model can be effective in analysis of other sports such as football, baseball, golf ball, and so on where accurate travel distance or even a long-distance travel caused by an impact should be measured.

## II. DEFINITION OF IMPULSE AND COR

When collision occurs, the amount of impulse during the contact interval is defined as

$$\underline{\tilde{F}}_{ext} = \int_{t_0}^{t_0+\Delta t} \underline{F}_{ext} dt. \tag{1}$$

In general, when the collision between two bodies is partially elastic, then the range of COR is $0 < e < 1$. If the COR $e$ is given, variation of the relative velocity along the normal vector $\underline{n}$, which is normal to the contact surface, can be derived as [30]

$$\left(\Delta v_M - \Delta v_B\right)^T \underline{n} = -(1+e)\left(v_M - v_B\right)^T \underline{n}, \tag{2}$$

where $v_M$ and $v_B$ are the absolute velocities of two bodies before the collision, and $\Delta v_M$ and $\Delta v_B$ are the velocity variation of the two bodies after the collision, respectively.

## III. AERODYNAMIC MODEL

The linear and angular velocities of a flying ball immediately after collision are decided by the direction and magnitude of the external impulse [6, 7] exerted to the ball. After the collision with another object, the ball will be influenced by not only gravity force but also two aerodynamic forces: the drag force and the force caused by Magnus effect. The drag force occurs when the resistance force between object and fluid exists. The magnitude of the drag force is known as

$$F_D = \rho S \frac{C_D v_{air}^2}{2}, \tag{3}$$

where $\rho$, $S$, $C_D$, and $v_{air}$ are the density of the air (1.226 kg/m³ in $15°C$, 1013 hpa), the cross-section area of the ball, coefficient of the drag force, and the relative velocity of the air with respect to the ball, respectively. The direction of the drag force is always the opposite to the progress direction of the ball as shown in Fig. 1.

Fig. 1

In Eq. (3), $\rho$ and $S$ are constant values. $C_D$ depends on the Reynolds number $N_{Red}$ which is the ratio of the inertial force to the viscous force. Reynolds number is expressed as follows

$$N_{Red} = \frac{\rho V D}{\mu}, \tag{4}$$

where $\rho$ is the density of the air mentioned before, $V$, $D$, and $\mu$ denote the average velocity of the air, the diameter of the ball, and the coefficient of the air viscosity ($1.83 \times 10^{-5}\ kg/m \cdot \sec$ in $15°C$). Therefore, the drag force is the function of $C_D$ and $v_{air}$ [31]. If there is no wind, $V = v_{air}$ and the drag coefficient $C_D$ can be derived by the plot which relates Reynolds number to $C_D$ shown in Fig. 2 [12, 31, 32].

Fig. 2

Meanwhile, the ball traveling in the air can also receive another force. If the ball rotates counter-clockwise, the air surrounding the ball also rotates counter-clockwise as shown Fig. 3. It is easily noted that in the upper part of the ball the air flow increases and the air pressure decreases by the Bernoulli's law. In the lower part of the ball, it is opposite. So, the air flow decreases and the air pressure increases. Because of the difference in the air pressure between the lower part and the upper part of the ball, there exists a lift force on the ball. This effect is referred as the "Magnus effect". The force caused by the Magnus effect makes the ball turn left or right and lift down or up according to the direction of the rotation axis of Fig. 4. In general 3 dimensional cases, the forces that make turn left or right and that make lift up or down occur, simultaneously.

Fig. 3

Fig. 4

The force caused by the Magnus effect is denoted as

$$F_{mag} = \rho S \frac{C_{mag} v_{air}^2}{2}, \tag{5}$$

where $C_{mag}$ is the coefficient caused by the Magnus effect. The direction is perpendicular to the linear velocity of the ball and the rotation axis. It is decided by taking cross product of the rotation axis and the ball velocity. In Eq. (5), $\rho$ and $S$ are constant values. $C_{mag}$ depends on the Reynolds number $N_{Red}$ in Eq. (4) and spin parameter (Sp) which is denoted as

$$Sp = \frac{wr}{v_{air}}, \tag{6}$$

where $w$, $r$, and $v_{air}$ are the angular velocity of the ball, the radius of the ball, and the velocity of the air, respectively. In the spin parameter which is presented in Eq. (6), $w$ and $v_{air}$ are the variables, and in the Reynolds number $N_{Red}$, $v_{air}$ is the variable. The coefficient of Magnus effect $C_{mag}$ can be analyzed by using the plot which relates spin parameter Sp to $C_{mag}$ ($C_L$) for corresponding Reynolds number which is presented as Fig. 5 [12, 31, 32].

Fig. 5

## IV. EXPERIMENTAL ENVIRONMENT

In this section, the experimental conception and the specific setting of experimental equipment are described. To introduce and analyze the variable characteristic of the COR model, the collision phenomenon between a pendulum and two kinds of ball is employed as an example. Specification for the elements of experimental environment and precautions are mentioned. The experimental environment is presented as Fig. 6.

Fig. 6

The experimenter can pull back the pendulum which is shown in Fig. 6 to a specific back swing angle. By locating the pendulum to arbitrary angular position, the potential energy is generated. When the pendulum is laid down, the potential energy is converted into the kinetic energy. Right after the collision of the pendulum on the ball, the impulse applied to the ball yields onset of the ball's motion, followed by the travel in the air. In this experiment, the equally-spaced squares are used to measure the travel distance of the ball more precisely, and the camcorder is also used to measure the contact position and behavior between the ball and the pendulum. Two kinds of ball, which are different in material and size, are employed in experiment.

Fig. 7

Fig. 8

The actual experimental environment is shown in Fig. 7, and its components are presented in Fig. 8. A pendulum with 1 DOF (degree of freedom) is introduced, which corresponds to a golf club, a bat in baseball, the human leg in jokku, and so on. 50mm-spaced squares and 10mm-spaced gradations are set on the wall as shown in Fig. 8(a) to measure the travel distance of the ball due to impact, and the protractor shown in Fig. 8 (b) is installed on the rotation axis of the pendulum to measure the back swing angle. In this experiment, a gadget for measuring the ball pressure is used to make the ball pressure retain consistently as depicted in Fig. 8 (c). To measure the angular position and velocity, an encoder is installed on the rotating axis of the pendulum. To gauge the contact time between the distal end of the pendulum and the ball, a FSR sensor is attached on the surface of the mass located at the distal end of the pendulum as shown in Fig. 8 (d) and (e), respectively. A microprocessor (MCU) shown in Fig. 8 (f) is used for measuring the digital signal which comes from the encoder and the FSR sensor. In order to prevent the delay caused by communication, both the encoder data and the FSR sensor data are stored to the microprocessor during a designated time, and then printing out the stored data on PC is followed. The specification for the setting of the experimental system is summarized in Table 1.

The inertia, mass, length data for the experiment setting are summarized in Table 2. The data for mass and length are measured by scale and ruler, and the inertia-related data is obtained by using the design software (Solidworks). The effective mass of the impacting pendulum at the position of impact is denoted as $M_E = Ml^2 + \frac{1}{3}ml^2$, where $M$, $m$, and $l$ are the mass attached to the distal end of the pendulum, the mass of the pendulum bar, and the length of the pendulum bar, respectively.

The ball's behavior after receiving the impulse is influenced by several environmental conditions. The environmental conditions of this experiment are set as Table 3. The temperature of the room is maintained consistently because the viscosity and density of the air are influenced by the temperature [26].

In this experiment, two kinds of ball (i.e., jokku ball and basketball) are used, and their properties are presented in Table 4. The contact angle implies the launch angle of the ball and it is measured at the contact moment. The ball pressure is measured by the gadget shown in Fig. 8 (c) and its value is retained consistently to make sure an identical experimental condition.

To prevent retroaction of the entire pendulum equipment, score of kilogram of the weight is laid on the supporter of the equipment. Moreover, it is noted that the snap shot image of the ball projected to the wall is larger than that of the original image. An error correction algorithm is implemented to supplement the distance difference caused by the projection as shown in Fig. 9.

Fig. 9

# V. ANALYTIC MOEDL OF COR AND ITS APPLICATION TO BALL KICKING PROBLEM

This section can be categorized into three parts. The first part is to derive the analytic model of COR (coefficient of restitution) based on the ball kicking experiment. The second part is to compare the travel distance between simulation and experiment for two kinds of ball. In the simulation, the travel distances for the curve-fitted constant COR model and the curve-fitted variable COR model are compared to verify the effectiveness of the analytical variable COR model. Furthermore, the proposed constant COR model and variable COR model are compared in several kinds of viewpoint. Using the curve-fitted variable COR model, the third part deals with estimation of the travel distance of the ball for bigger collision velocity. We claim that the travel distance is considerably affected by the COR. Then, the proposed results can be beneficially applied to many sports field such as basketball, soccer, baseball, jokku, golf, and so on.

## 1. Analytic modeling of COR through two kinds of ball kicking experiment

If the ball receives an impact by the impacting pendulum, an external impulse $\left(\tilde{F}_{ext}\right)$ is exerted on the ball and the impacting pendulum also receives the same external impulse in the opposite direction as shown in Fig. 10.

Fig. 10

The external impulses exerted on the ball and the impacting pendulum are derived as, respectively

$$\tilde{F}_{ext} = m_B \Delta \underline{v}_B = m_B(\underline{v}_B - \underline{v}_B^0), \tag{7}$$

$$-\tilde{F}_{ext} = M_E \Delta \underline{v}_M = M_E (\underline{v}_M - \underline{v}_M^0), \tag{8}$$

where $m_B, M_E, \underline{v}_B^0, \underline{v}_B, \underline{v}_M^0$, and $\underline{v}_M$ are the mass of the ball, the effective mass at the distal end of the impacting pendulum, the velocity of the ball before and after the impact, the velocity of the pendulum before and after the impact, respectively. By rearranging Eq. (7) and (8) with respect to the velocity variation and then substituting into Eq. (2), the external impulse exerted on the ball can be derived as the following equation

$$\tilde{F}_{ext} = \frac{M_E m_B (1+e) \underline{v}_M^0}{M_E + m_B}, \tag{9}$$

where it is noted that the initial ball velocity $\underline{v}_B^0 = 0$.

The velocity of the ball immediately after the impact can be solved by substituting Eq. (9) into Eq. (7) and then rearranging with respect to $\underline{v}_B$ as follows

$$\underline{v}_B = \frac{M_E (1+e) \underline{v}_M^0}{M_E + m_B}. \tag{10}$$

Meanwhile, the force caused by the Magnus effect is ignored, because the impacting pendulum hits toward the center of mass of the ball and thus there is no rotational movement of the ball in this experiment. Therefore, the dynamic equations of the ball in x- and y-directions are presented as follows

$$m_B \ddot{x} = -(F_D)_x, \tag{11}$$

$$m_B \ddot{y} = -m_B g \pm (F_D)_y, \tag{12}$$

The meaning of the plus or minus sign conversion in Eq. (12) is that the direction of the drag force in y-component can be changed according to the direction of the ball velocity. The tendency of the gravitational force and the drag force for the flying ball is depicted in Fig. 11.

Fig. 11

The "$x$" coordinate of the traveling ball is derived as

$$x = \frac{m_B}{\frac{1}{2}\rho S C_D} \ln\left(1 + \frac{\frac{1}{2}\rho S C_D (v_B \cos\theta)}{m_B} t\right) + \left(v_M^{avg} \cos\theta\right) \Delta t \tag{13}$$

by integrating Eq. (11) with respect to time twice. The second term is added since the pendulum and the ball moves together during the contact time $\Delta t$. $\theta$ denotes the launch angle of two kinds of the ball. $v_M^{avg} = \frac{v_M^0 + v_M}{2}$ implies the average velocity of the pendulum during the period of impact. In this experiment, the velocity component in the y-direction (vertical component) is much smaller than that in the x-direction. Therefore, the drag force in the y-direction is negligible and only gravitational force has an effect on the motion in the y-direction. Given the launching angle of the ball, the "y" coordinate of the traveling ball is derived as

$$y = (v_B \sin\theta)t - 0.5gt^2 + \left(v_M^{avg} \sin\theta\right)\Delta t, \tag{14}$$

by integrating Eq. (12) with respect to time twice. The third term is also added since the pendulum and the ball moves together during the contact time. The travel time $t$ is defined as the time duration from the launching time to the ground arriving time of the ball. It can be derived by solving Eq. (14) with respect to $t$, and introducing the height of the supporter $h_s$ as follows

$$t = \frac{v_B \sin\theta + \sqrt{(v_B \sin\theta)^2 + 4 \cdot 0.5g \cdot \left(v_M^{avg} \sin\theta\right)\Delta t + h_s}}{2 \cdot 0.5g}. \tag{15}$$

Substituting Eq. (10) and Eq. (15) into Eq. (13), the "$x$" coordinate of the traveling ball can be rewritten as

$$x = \frac{m_B}{\frac{1}{2}\rho S C_D} \ln\left(1 + \frac{\rho S C_D M_E (1+e) \underline{v}_M^0 \cos\theta}{2 m_B (M_E + m_B)} \cdot \frac{v_B \sin\theta + \sqrt{(v_B \sin\theta)^2 + 2g(v_M^{avg} \sin\theta \cdot \Delta t + h_s)}}{2 \cdot 0.5 g}\right)\ldots \quad (16)$$
$$+ (v_M^{avg} \cos\theta) \Delta t,$$

and rearranging Eq. (16), the analytical model of the COR($e$) can be obtained as

$$e = \left[\frac{M_E + m_B}{M_E v_M^0} \cdot \frac{\frac{g}{K}\left(\exp^{K(x - v_M^{avg} \cos\theta \Delta t)} - 1\right)}{\sqrt{\sin 2\theta \cdot \frac{g}{K}\left(\exp^{K(x - v_M^{avg} \cos\theta \Delta t)} - 1\right) + 2g(\cos\theta)^2 \cdot (v_M^{avg} \sin\theta \cdot \Delta t + h_s)}}\right] - 1 \quad (17)$$

It should be noted that "e" (COR) is functions of many properties; the contact time $\Delta t$, the velocity $v_M^0$ of the pendulum before impact, the drag coefficient $C_D$, masses of the system the cross-sectional area of the ball, and the density of the air. Analyzing the effect of each parameter on COR, we come to some conclusions as follow through numerical analysis.

1. As the colliding velocity of the pendulum increases, COR gets decreased.

2. As the contact time increases, COR gets decreased.

3. As mass of the ball increases, COR gets increased.

4. The cross sectional area of the ball, air density, or drag coefficient do not affect COR much.

In (17), $K = \rho S C_D / 2 m_B$. It is remarked that this COR model is more general as compared to the variable COR models of previous works.

In Eq. (17), $y$ and masses are known from Table 2. The launching angle is given as $\theta = 8.5$

and 11.8 (deg) for jokku ball and basketball, respectively, the drag coefficient is obtained from Fig. 2, $v_M^0$ and $x$ are measured by an encoder mounted at the rotating axis of the pendulum and a camcorder fixed on the ground, respectively. The experiment is conducted by measuring the distance for several back swing angles such as 30, 60, 90, and 120 (deg). Performing more than 20 experiments for each of 4 back swing angles, the tendencies of the travel distance with respect to the back swing angle for jokku ball and basketball are shown as Fig. 12.

Fig. 12

The contact time during the collision between the pendulum and the ball is measured by FSR sensor. However, to verify the performance of FSR sensor, the contact time is also measured by a camcorder by shooting 600 frames per a second for each back swing angle. The number of the snapshot is counted more than 10 times and averaged. The tendencies of the contact time with respect to the back swing angle for jokku ball and basketball are shown as Fig. 13. Though the material properties of the two balls are different, the tendencies of the contact time measured by the FSR sensor and the camcorder were similar to each other. Thus, we can confirm the accuracy of the FSR.

Fig. 13

In this experiment, the difference of the potential energy between the initial and final positions of the pendulum is taken into account to measure the initial contact velocity $v_M^0$ at the distal end of the pendulum. The concept of calculating the energy loss is depicted as Fig. 14.

Fig. 14

The amount of the energy loss is derived as

$$E_{loss} = M_E gL(1-\cos\theta_1) - M_E gL(1-\cos\theta_2) = M_E gL(\cos\theta_2 - \cos\theta_1) = \frac{1}{2}M_E\left(w_1^2 - w_2^2\right) \quad (18)$$

where $\theta_1$, $\theta_2$, $w_1$, $w_2$ are the back swing angle, the forward swing angle after the impact ($\theta_1 > \theta_2$), the angular velocity of the pendulum immediately before and after the impact, respectively. It is assumed that the energy loss is caused by not only the impact, but also the friction at the rotating axis. To make the above method reasonable, a smooth bearing is installed at the axis of the pendulum to reduce the friction torque and the retroaction of the pendulum supporter by loading the weight on the hardware. Since the angular velocities of the impacting pendulum can be written in terms of velocity of the pendulum such as $w_1 = v_M^0/L_E$ and $w_2 = v_M/L_E$, Eq. (18) can be rewritten as

$$E_{loss} = \frac{1}{2}M_E\left(w_1^2 - w_2^2\right) = \frac{1}{2}\frac{M_E}{L_E^2}\left(v_M^{0\,2} - v_M^{\,2}\right) = \frac{1}{2}\frac{M_E}{L_E^2}\left(v_M^0 + v_M\right)\left(v_M^0 - v_M\right). \quad (19)$$

In this experiment, the term $v_M^0 + v_M$ is approximated as $v_M^0 + v_M \simeq 2v_M^0$ and the term $v_M^0 - v_M$ given by Eq. (8) is used. Then, the energy loss is expressed as

$$E_{loss} = \frac{1}{2}\frac{M_E}{L_E^2}\left(v_M^0 + v_M\right)\left(v_M^0 - v_M\right) = \frac{1}{2}\frac{M_E}{L_E^2}2v_M^0 \frac{\tilde{F}_{ext}}{M_E}. \quad (20)$$

Substituting Eq. (9) into Eq. (20), the following equation is derived.

$$E_{loss} = \frac{1}{2}\frac{M_E}{L_E^2}2v_M^0 \frac{m_B(1+e)v_M^0}{M_E + m_B}. \quad (21)$$

Rearranging Eq. (17), $v_M^0$ can be expressed as the follows

$$v_M^0 = \left(\frac{M_E + m_B}{M_E(e+1)}\right) \cdot \left(\frac{\frac{g}{K}\left(\exp^{Kx-\left(v_M^{avg}\cos\theta\right)\Delta t} - 1\right)}{\sqrt{\sin 2\theta \cdot \frac{g}{K}\left(\exp^{Kx-\left(v_M^{avg}\cos\theta\right)\Delta t} - 1\right) - 2g(\cos\theta)^2 \cdot \left(y - \left(v_M^{avg}\sin\theta\right)\Delta t\right)}}\right), \quad (22)$$

and substituting Eq. (22) into Eq. (21), the energy loss is derived as

$$E_{loss} = \frac{1}{L_E^2} \cdot \frac{m_B(M_E + m_B)}{M_E(e+1)} \cdot \left(\frac{\frac{g}{K}\left(\exp^{Kx-\left(v_M^{avg}\cos\theta\right)\Delta t} - 1\right)}{\sqrt{\sin 2\theta \cdot \frac{g}{K}\left(\exp^{Kx-\left(v_M^{avg}\cos\theta\right)\Delta t} - 1\right) - 2g(\cos\theta)^2 \cdot \left(y - \left(v_M^{avg}\sin\theta\right)\Delta t\right)}}\right)^2. \quad (23)$$

In Eq. (23), it is observed that the quantity of the energy loss increases as COR(e) value decreases and that a large back swing angle increases the energy loss. This tendency is observed in Fig. 15. To consider only the energy loss caused by the impact, it is obtained by subtracting the energy loss with no impact (black dotted-line) from the total energy loss (black solid-line) and it is presented as red dotted-line in Fig. 15 for jokku ball and basketball, respectively.

Fig. 15

The linear velocity $v_M^0$ of the pendulum before impact is derived from angular velocity of the pendulum which is measured by the encoder. The linear velocity $v_M$ of the pendulum after impact is derived by subtracting energy loss quantity which is presented as red dotted-line in Fig. 15. The tendencies of linear velocity of the pendulum before and after the impact for two different balls are plotted as Fig. 16. It is noted that the magnitude of the linear velocity after the impact is less than before the impact due to the energy loss.

Fig. 16

In this experiment, it is usually hard to measure the external impulse experimentally because the period of impact is too short. So we employ Eq. (8) to calculate the external impulse indirectly since the effective mass is known. Its result is shown as Fig. 17. The initial velocities $v_b^0$ of the balls after the impact with respect to the back swing angle are solved by substituting the mass of each ball into Eq. (7). To verify this result, the initial velocities of two balls after the impact are measured by camcorder. The displacement of the ball during unit frame (600 frames per second) of camcorder is used for measuring the ball velocity. Figure 18 shows that the velocity based on Eq. (7) (black solid-line) is compared with the data measured by the camcorder (red dotted-line). It is noted that the velocity profiles for two cases are almost identical.

Fig. 17

Fig. 18

Finally, using Eq. (9), CORs of two different balls with respect to the back swing angle are depicted in Fig. 19. In Fig. 20, CORs of two different balls with respect to the linear velocity of the pendulum before the impact are shown. It is noted that the COR decreases as the back swing angle or the pendulum's velocity increases, which is coincident to result of many previous researches [9-29].

Fig. 19

Fig. 20

*2. Comparison between constant COR and the analytically-obtained variable COR*

In this section, the effectiveness of the variable COR is presented by observing the difference between constant COR and variable COR based on the results of the impact experiment. Three comparison criteria between constant COR and variable COR are employed: difference of travel distance, external impulse for the same collision velocity, and difference of back swing angle to yield the same external impulse.

First, it is proved that the variable COR is more applicable to the impact phenomenon than the constant COR by comparing the simulated travel distance to the actual travel distance achieved by the experiment. Before comparison, a curve-fitted variable COR model was developed to set the constant value of the constant COR model. To see the tendency of variable COR, refer to the paper presented by Andersen [11], which conducts experiment with large contact velocity. The constant value of the constant COR model is set by the value at the point of $\underline{v}_M^0 = 0$ of the curve-fitted variable COR model, since there exists no energy loss when the collision velocity is zero. The curve fitting form of variable COR for jokku ball and basketball are selected as a negative exponential function. The curve-fitted model of constant COR and variable COR for jokku ball are presented in Eq. (24), and for basketball in Eq. (25). The trends for Eq. (24) and Eq. (25) are plotted in Fig. 21. Using this model, the COR value for arbitrary $\underline{v}_M^0$ can be estimated.

$$\begin{aligned} \text{variable COR jokku}(\underline{v}_M^0) &= 0.1305 e^{-0.2685 \underline{v}_M^0 - 0.1953} + 0.6746 \\ \text{constant COR jokku}(\underline{v}_M^0) &= 0.1305 e^{-0.1953} + 0.6746 \end{aligned} \qquad (24)$$

$$\text{variable COR basket}(v_M^0) = 0.2235 e^{-0.2221 v_M^0 - 0.7226} + 0.8327$$
$$\text{constant COR basket}(v_M^0) = 0.2235 e^{-0.7226} + 0.8327$$

(25)

Fig. 21

Variable COR affects the velocity and drag coefficient after impact. Therefore, there exists some difference in travel distance between the variable COR and constant COR. The simulation to observe the travel distance of the ball is constructed as the following process.

a. For the given $\underline{v}_M^0$, solve the velocity $\underline{v}_B$ of the ball after impact by using the curve-fitted COR given in Eq. (24) or (25).

b. Choose the drag coefficient from Fig. 2 and calculate the drag force from Eq. (3).

c. Integrate the differential equations presented in Eq. (11) and (12) with respect to time, and solve the velocity of the ball $\underline{v}_B$ in the next step by using 4-th order Runge-Kutta method.

d. Integrate the differential equations one more time at "step c" by using 4-th order Runge-Kutta method to obtain the x- and y-coordinates of the jokku or basketball.

e. Iterate step from "b" to "d" until the y coordinates is near to zero (i.e., landing on the ground).

Using the above simulation process, the travel tendency of two kinds of ball for each back swing angle is plotted as Fig. 22. In Fig. 22, the solid-lines denote the simulated travel tendency when the variable COR is applied, and the dotted-lines denote the travel distances for the constant COR. 4 dots which are in the x-axis denote the experimentally measured average travel distances corresponding to each back swing angle. The entire statistics for the travel distance is summarized in Table 5.

Fig. 22

It is shown from Fig. 22 and Table 5 that the travel distances for both constant COR (A) and variable COR (B) increase as the back swing angle of pendulum increases. Moreover, the error between constant COR and experiment increases as the back swing angle increases ( $((B-C)/C) \cdot 100(\%)$ ), while the error between the variable COR and experiment is consistently smaller than 1% even though the back swing angle increases ($((A-C)/C) \cdot 100(\%)$). It can be inferred from the results of Fig. 22 and Table 5 that the estimated travel distance based on variable COR is more accurate than the estimated travel distance based on constant COR.

Secondly, comparison of the external impulse quantity between constant COR and variable COR for the same collision velocity of the pendulum is carried out. For two types of balls, external impulses for variable COR and constant COR can be calculated by substituting Eq. (24) and Eq. (25) into Eq. (9). Its tendency and specific values are presented in Fig. 23 and Fig. 24, and Table 6, respectively. It is shown that the difference in external impulse between constant COR and variable COR increases as the collision velocity of the pendulum increases. When the back swing angle is 120 (deg), the difference of the external impulse for jokku ball and basketball is about 0.09Ns, and 0.13Ns, respectively.

Fig. 23

Fig. 24

Moreover, we analyze how much collision velocity is needed to achieve the same external impulse for constant COR and variable COR. The collision velocity depicted in Fig. 25 and Fig. 26, and its numerical values written in Table 7 was derived by substituting Eq. (24) and Eq. (25) into Eq. (9) and arranging it with respect to $\underline{v}_M^0$.

Fig. 26

Fig. 27

It can be inferred that variable COR needs larger pendulum collision velocity than constant COR to produce the same external impulse because of the decrease in COR caused by energy loss. Its gap gets larger as the external impulse increases and there exists about 0.12m/s gap in 3.8Ns external impulse.

Finally, the difference of the back swing angle between constant COR and variable COR to exert the same external impulse is considered. Before that, the curve fitted model of the linear velocity of the pendulum before impact should be obtained with respect to the back swing angle. The curve fitted model of linear pendulum velocity before impact ($\underline{v}_M^0$) with respect to the back swing angle (b.s.a.) is constructed to estimate the travel distance of the ball by using back swing angle. Curves are set as a rational function. Corresponding equations and its trends are given in Fig. 27, and Eq. (26) and (27), respectively.

Fig. 27

$$\text{linear pen. vel. bef. imp jokku} = \frac{9.594 \cdot \text{b.s.a.}^2 - 85.5 \cdot \text{b.s.a.} - 0.2311}{\text{b.s.a.}^2 + 152.3 \cdot \text{b.s.a.} + 6.66} \tag{26}$$

$$\text{linear pen. vel. bef. imp basket} = \frac{9.554 \cdot \text{b.s.a.}^2 - 98.12 \cdot \text{b.s.a.} - 1.5070}{\text{b.s.a.}^2 + 154.2 \cdot \text{b.s.a.} + 7.267} \tag{27}$$

where b.s.a. implies the back swing angle. Then, substituting Eq. (26) and (27) into Eq. (24) and Eq.(25), respectively, the tendencies and specific values of the back swing angle with respect to the external impulse can be observed as Fig. 28 and Fig. 29, and Table 8. From the figures and tables, it is inferred that variable COR needs more external impulse than constant COR to achieve the same external impulse in every range, and its gap gets larger as the external impulse increases. As shown in Table 8, there exist about 6.5 (deg) and 5.0 (deg) difference in 2.2Ns and 3.8Ns for jokku ball and basketball, respectively.

Fig. 28

Fig. 29

## *3. Application to making the ball reach to the desired distance*

In this section, comparison of the travel distance between variable COR-based simulations and experiment is conducted for an exceeded range of collision velocity caused by more than 120 (deg) back swing angle for two kinds of ball. Using the previously used simulation which is presented in Fig. 22, the back swing angle of the pendulum, which corresponds to the desired travel distance, is written as Table 9.

For both jokku ball and basketball, four desired travel distances less than 2.2(m) in both are derived from the curve fitted variable COR model within the 120(deg) back swing angle using the experimental results, and the travel distance for more than 120(deg) back swing angle can be also estimated by the curve fitted variable COR model. Given as Table 9, the experiment is repeatedly conducted, and its experimental tendency and values are presented in Fig. 30 and Fig. 31, and Table 10, respectively. It is observed that the error between experimental travel distance denoted as (A) and simulated travel distance denoted as (B) is less than 1% for each ball ( $((A-B)/B)\cdot 100(\%)$ ). Therefore, it is remarked that in both balls, the proposed curve-fitted variable COR model is also applicable to more than 120 (deg) back swing angle or general collision velocity.

Fig. 30

Fig. 31

## V. CONCLUSION

Previous researches related to variable characteristic of COR are mostly based on experimental observation or numerical method. The main contribution of this paper is a suggestion of an analytical COR model which is more general as compared to previous researches. Two kinds of the ball with different physical properties were employed for experimental verification of the COR model. Next, using this analytical COR model and aerodynamic model, we found that considerable difference is found in travel distance of two different balls between variable COR model and constant COR model through simulation and that variable COR model is more precise than constant COR by comparing simulation results and experimental results. As the velocity of the impacting pendulum increases, the difference in travel distance becomes increased. Therefore, the introduction of variable COR model is important for precisely controlling the travel distance of the ball. Finally, comparison of travel distance between proposed variable COR model and experiment is conducted to verify this curve-fitted modeling even for large collision velocity. The proposed impact control can be beneficially applied to motion planning and estimation in many sports actions such as basketball, baseball, soccer, golf, and so on, where the ball is desired to reach by some distance. Moreover, not only in sports area but also in some profound physics fields, this research can be extended.

**REREFENCES**

Table 1. Specification of experimental setting

| Setting elements | Model | Specification |
|---|---|---|
| Encoder sensor | USdigital | 4096 pulses/turn |
| FSR sensor | FlexiForce® | 10us response time |
| MCU | dsPIC33FJ128MC802 | 10MHx clock |
| Protractor | printing | - 5(deg) resolution<br>- Maximum 140(deg) back swing angle |
| Ball pressure measurement | STAR digigauge | 0.05 psi resolution |
| Camcoder | Samsung HMX-S10BD | 300/600 frame shooting |

Table 2. Specification for length, mass and inertia properties

| Setting elements | Specification |
|---|---|
| Mass of the distal end | 4.750 kg |
| Mass of the bar | 1.250 kg |
| Total mass | 6.0 kg |
| Effective mass at the contact point | 5.1667 kg |
| COM | 0.4553 m |
| Moment of inertia of the pendulum about the rotating axis | 1.579 $kg \cdot m^2$ |
| Pendulum length | 0.5541 m |
| Height of the ball support | 0.3120 m |

Table 3. Specification of experiment for spatial setting

| Environment elements | Specification |
|---|---|
| Gravitational acceleration | 9.806 $m/s^2$ |
| Temperature | 26 Celsius degree |
| Air density | 1.180 $kg/m^3$ |
| Air dynamic viscosity | 1.852e-5 $N \cdot s/m^2$ |

Table 4. Specification for the two balls

|  | **Jokku ball specification** | **Basketball specification** |
|---|---|---|
| Model | Nassau ZEPPE 5.0 | Sparding Neverflat |
| Mass | 0.353 kg | 0.605 kg |
| Diameter | 201.75 mm | 245.90 mm |
| Contact angle | 8.5 degree | 11.8 degree |
| Ball pressure | 4.60psi | 14.45psi |

Table 5. Comparison of the travel distance with simulation and experiment for two kinds of ball

| Back sw. ang.(deg) / **Jokku** trav. Dist.(m) | 30 deg | 60 deg | 90 deg | 120 deg |
|---|---|---|---|---|
| Estimated by var. COR – (A) | 0.5148 | 1.198 | 1.808 | 2.341 |
| Estimated by const. COR – (B) | 0.5234 | 1.236 | 1.881 | 2.481 |
| Experiment (average) – (C) | 0.5181 | 1.1893 | 1.7932 | 2.3603 |
| (A-C)/C (%) | -0.6409 % | 0.7348 % | 0.8235 % | -0.8196 % |
| (B-C)/C (%) | 1.0189 % | 3.9300 % | 4.8943 % | 5.1118 % |
| Back sw. ang.(deg) / **Basket** trav. Dist.(m) | 30 deg | 60 deg | 90 deg | 120 deg |
| Estimated by var. COR – (A) | 0.5701 | 1.339 | 2.063 | 2.732 |
| Estimated by const. COR – (B) | 0.5764 | 1.375 | 2.139 | 2.861 |
| Experiment (average) – (C) | 0.5688 | 1.3354 | 2.0716 | 2.7500 |
| (A-C)/C (%) | 0.2219 % | 0.2731 % | -0.4149 % | -0.6555 % |
| (B-C)/C (%) | 1.3294 % | 2.9690 % | 3.2538 % | 4.0353 % |

Table 6. Comparison of external impulse with respect to the collision velocity for two kinds of ball

| Jokku Ext. impulse (Ns) \ Back sw. ang.(deg) | 30 deg | 60 deg | 90 deg | 120 deg |
|---|---|---|---|---|
| Var. COR (N*s) | 0.6510 | 1.3197 | 1.8267 | 2.2155 |
| Const. COR (N*s) | 0.6614 | 1.3574 | 1.8926 | 2.3059 |

| Basket Ext. impulse (Ns) \ Back sw. ang.(deg) | 30 deg | 60 deg | 90 deg | 120 deg |
|---|---|---|---|---|
| Var. COR (N*s) | 1.0771 | 2.2774 | 3.1908 | 3.8927 |
| Const. COR (N*s) | 1.0896 | 2.3280 | 3.2826 | 4.0213 |

Table 7. Comparison of collision velocity with respect to the external impulse for two kinds of ball

| Jokku Colliding vel. (m/s) \ Ext. impulse.(N*s) | 30 deg | 60 deg | 90 deg | 120 deg |
|---|---|---|---|---|
| Var. COR (m/s) | 1.740 | 2.449 | 3.166 | 3.888 |
| Const. COR (m/s) | 1.698 | 2.378 | 3.057 | 3.736 |

| Basket Colliding vel. (m/s) \ Ext. impulse.(N*s) | 30 deg | 60 deg | 90 deg | 120 deg |
|---|---|---|---|---|
| Var. COR (m/s) | 1.451 | 2.436 | 3.432 | 3.732 |
| Const. COR (m/s) | 1.427 | 2.370 | 3.329 | 3.614 |

Table 8. Comparison of back swing angle with respect to the external impulse for two kinds of ball

| Jokku Back sw. ang. (deg) \ Ext. impulse.(N*s) | 1.0 Ns | 1.4 Ns | 1.8 Ns | 2.2 Ns |
|---|---|---|---|---|
| Var. COR (deg) | 45.78 | 64.73 | 88.58 | 118.95 |
| Const. COR (deg) | 44.73 | 62.35 | 84.94 | 112.47 |
| Basket Back sw. ang. (deg) \ Ext. impulse.(N*s) | 1.5 Ns | 2.5 Ns | 3.5 Ns | 3.8 Ns |
| Var. COR (deg) | 40.76 | 67.31 | 103.38 | 116.19 |
| Const. COR (deg) | 40.14 | 65.39 | 99.00 | 111.17 |

Table 9. Back swing angle corresponding to the desired travel distance – jokku ball

| Desired travel distance (m) | 1.0 m | 1.4 m | 1.8 m | 2.2 m | 2.5 m |
|---|---|---|---|---|---|
| Jokku Corresponding back sw. ang. (deg) | 52.64 | 69.16 | 89.36 | 111.62 | **129.81** |
| Desired travel distance (m) | 1.0 m | 1.4 m | 1.8 m | 2.2 m | 2.8 m |
| Basket Corresponding back sw. ang. (deg) | 49.01 | 65.05 | 81.88 | 99.61 | **128.90** |

Table 10. Comparison of travel distance between desired and experiment for two kinds of ball

| Desired dist..(m) / Jokku trav. Dist.(m) | 1.0 m | 1.4 m | 1.8 m | 2.2 m | 2.5 m |
|---|---|---|---|---|---|
| Experiment in average – (A) | 1.0024 | 1.4055 | 1.8068 | 2.2082 | 2.5183 |
| Desired – (B) | 1.0 | 1.4 | 1.8 | 2.2 | 2.5 |
| (A-B)/B (%) | 0.2432 % | 0.3569 % | 0.3765 % | 0.3704 % | 0.7319 % |
| Back sw. ang.(deg) / Basket trav. Dist.(m) | 1.0 m | 1.4 m | 1.8 m | 2.2 m | 2.8 m |
| Experiment in average – (A) | 1.0029 | 1.4067 | 1.8076 | 2.2096 | 2.8178 |
| Desired – (B) | 1.0 | 1.4 | 1.8 | 2.2 | 2.8 |
| (A-B)/B (%) | 0.2852 % | 0.4765 % | 0.4247 % | 0.4376 % | 0.6362 % |

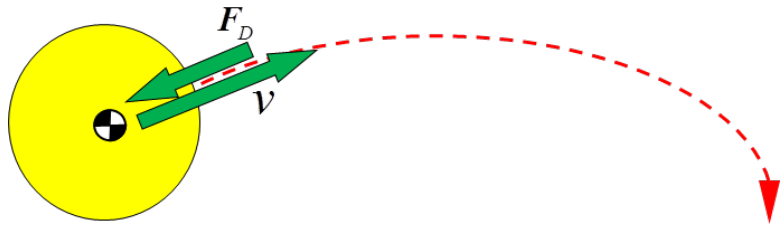

Fig. 1. Drag force for the ball flying in the air

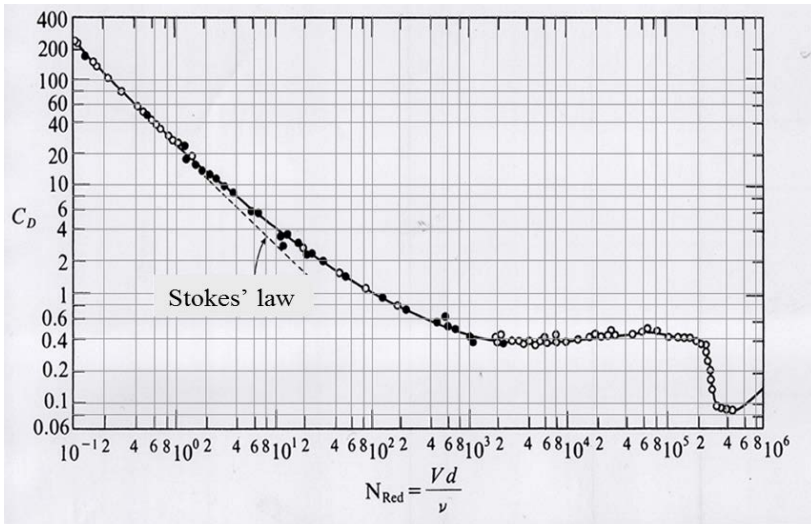

Fig. 2. The drag coefficient vs Reynolds number [12, 31, 32]

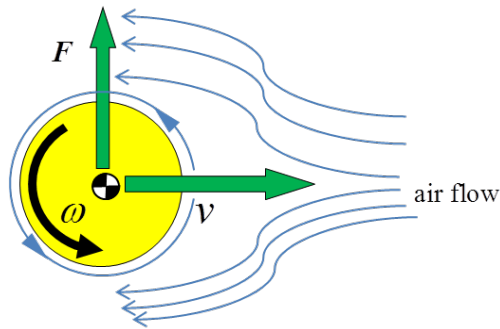

Fig. 3. Drag force for the ball flying in the air

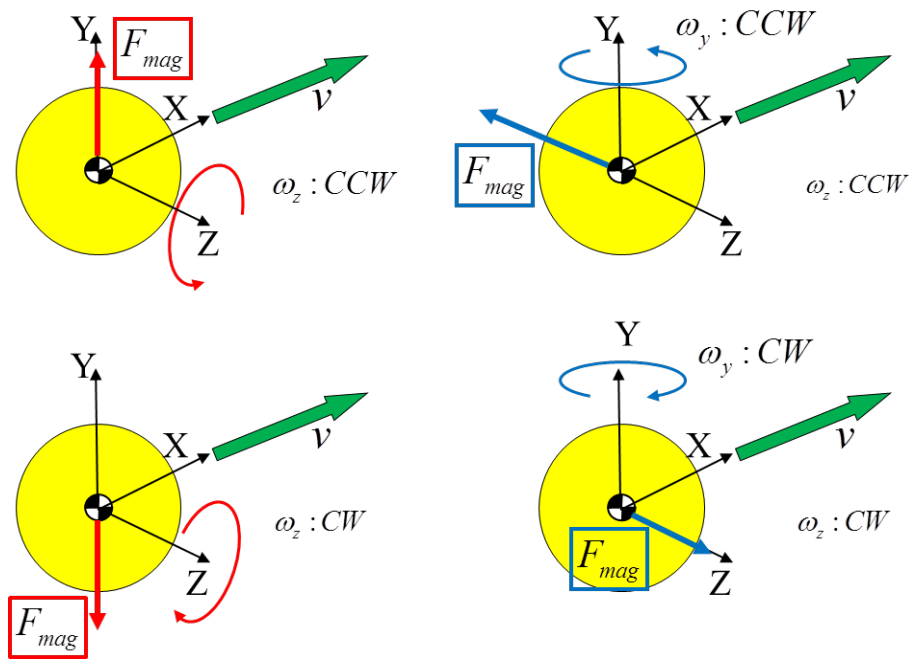

Fig. 4. 4 cases of the force caused by the Magnus effect

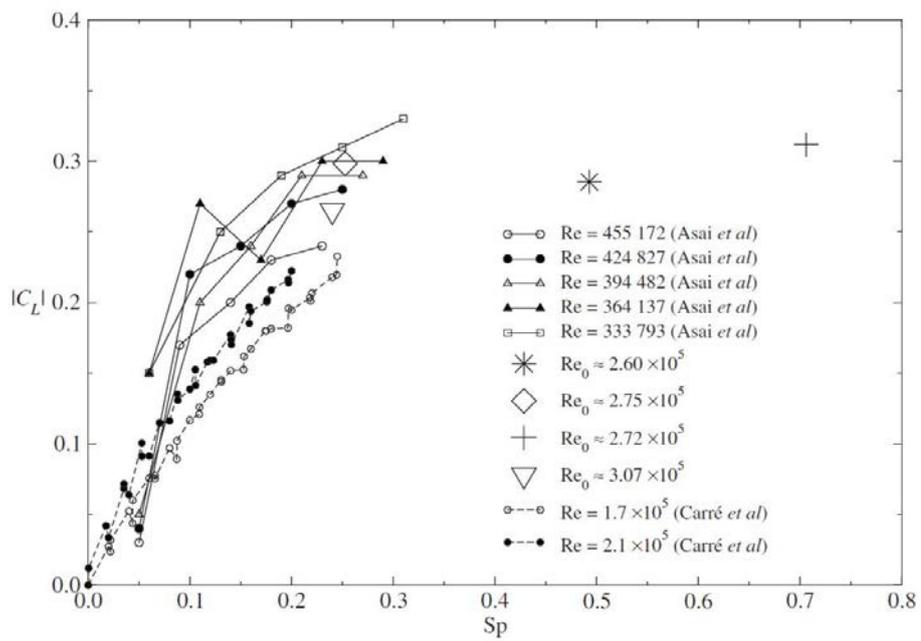

Fig. 5. The coefficient of the Magnus effect plot [12, 31, 32]

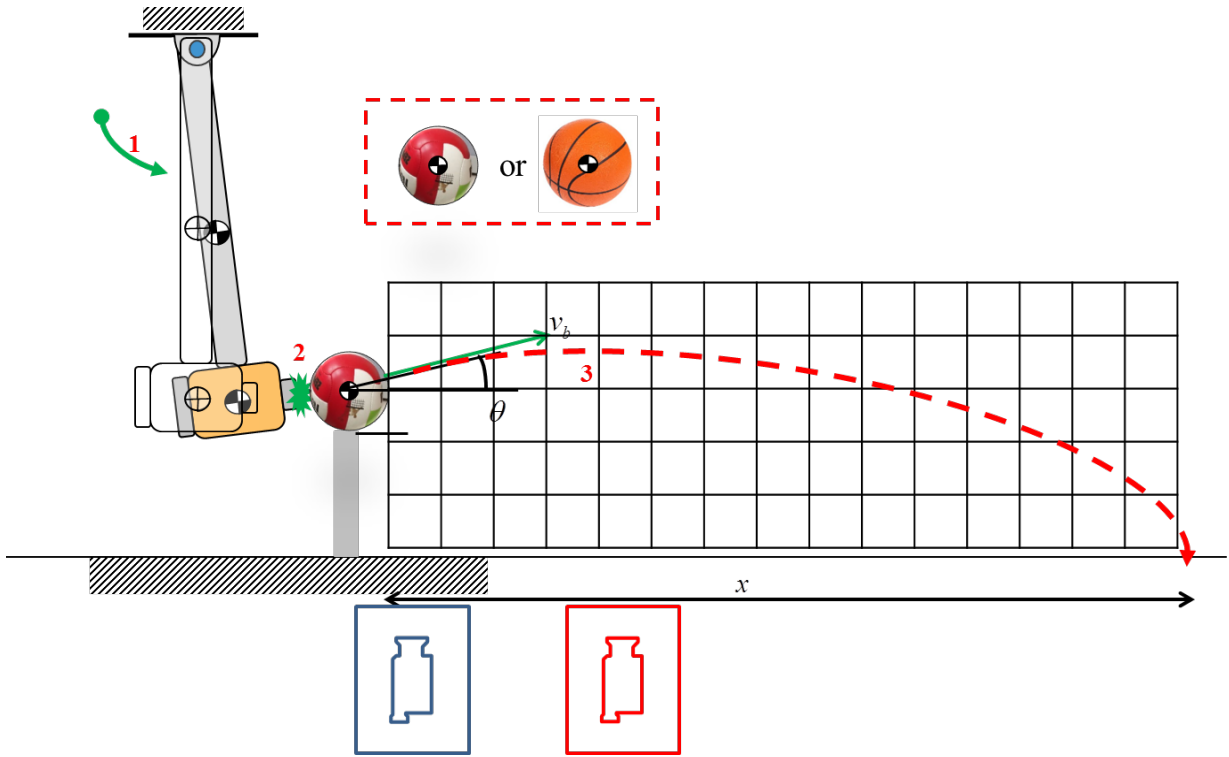

Fig. 6. Setup of the ball kicking experiment

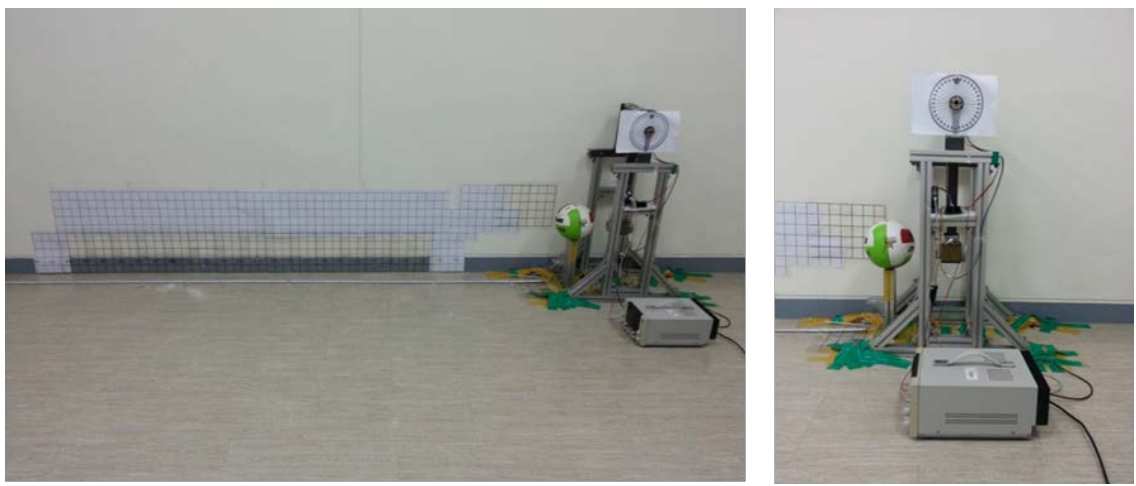

Fig. 7. Panorama of the experimental environment

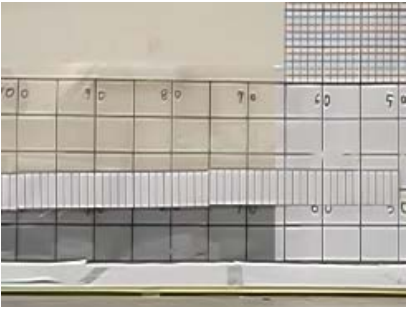 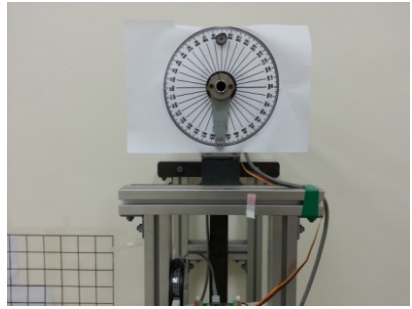 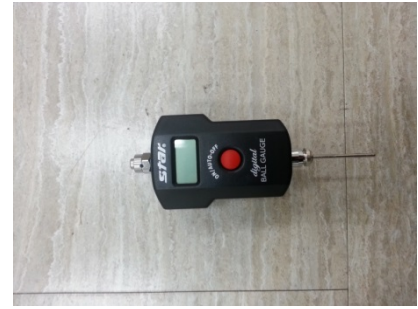

(a) 50mm and 10mm gradations     (b) Protractor     (c) ball pressure measurement gear

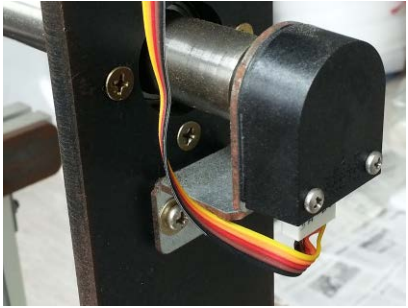 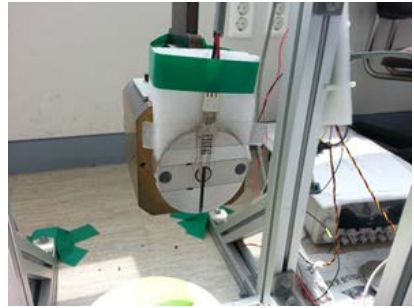 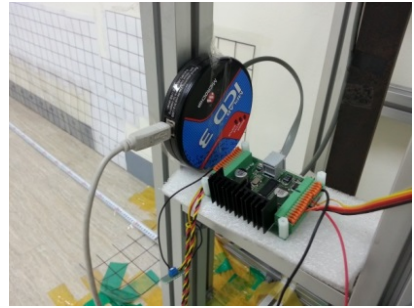

(d) Encoder     (e) FSR sensor     (f) MCU

Fig. 8. Elements of experiment environment

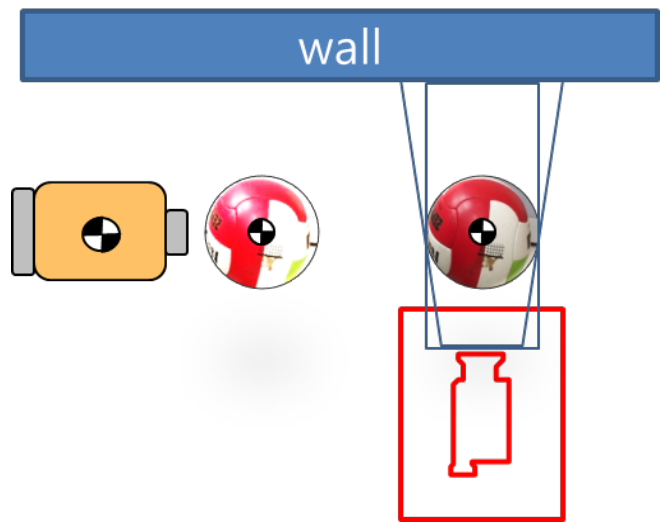

Fig. 9. Projection of the ball image to the wall

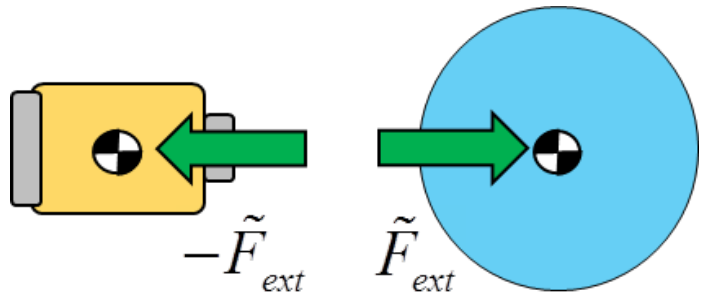

Fig. 10. Action-reaction law in the collision situation

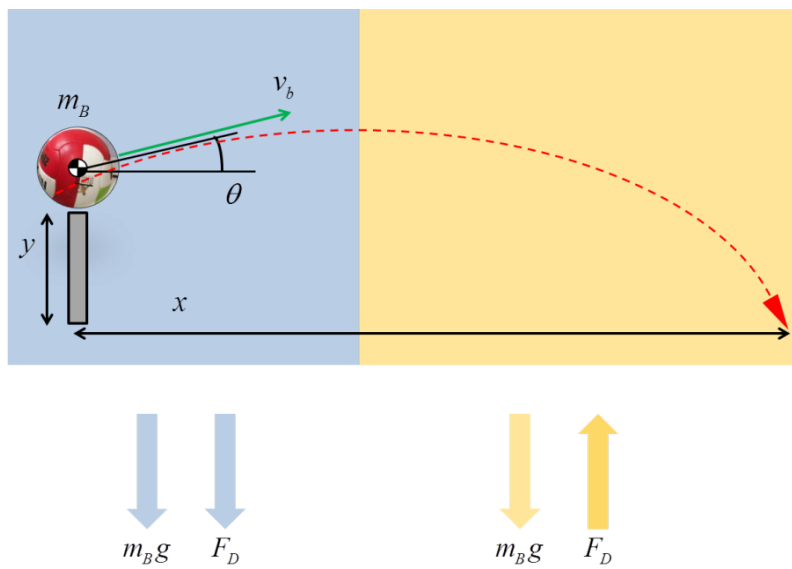

Fig. 11. Tendency of the gravity force and the drag force for the flying ball

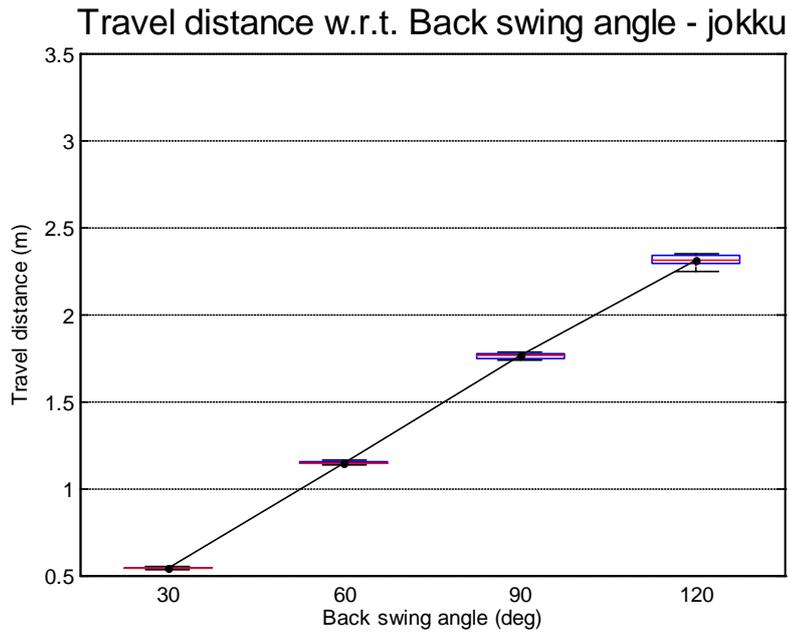
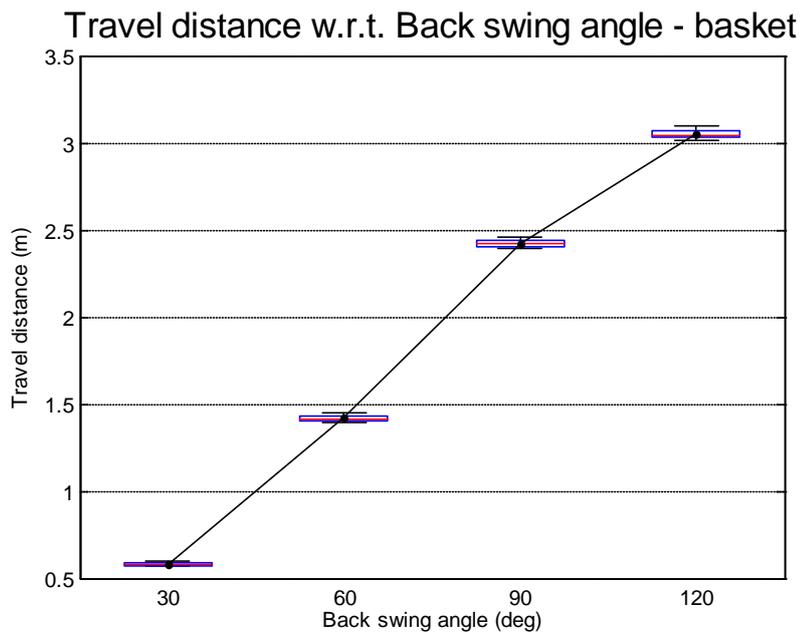

Fig. 12. Travel distance tendency with respect to back swing angle

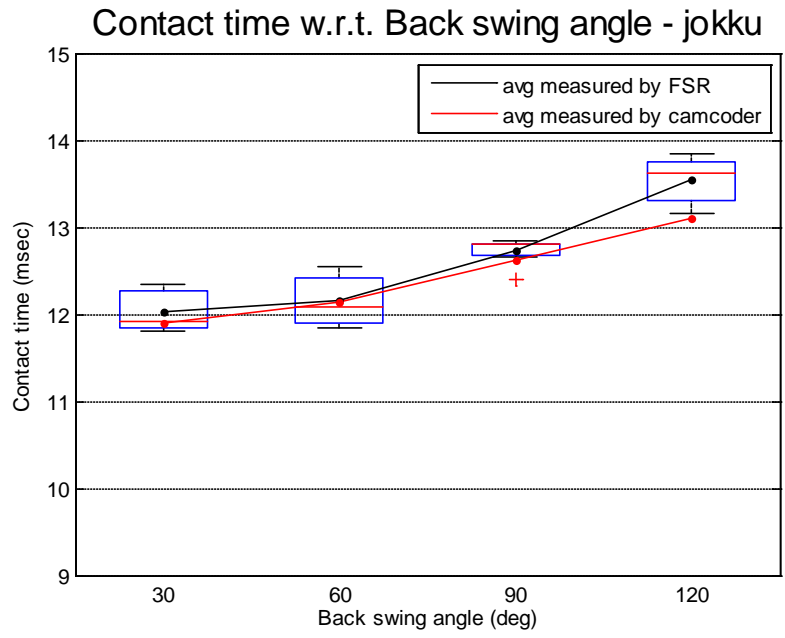

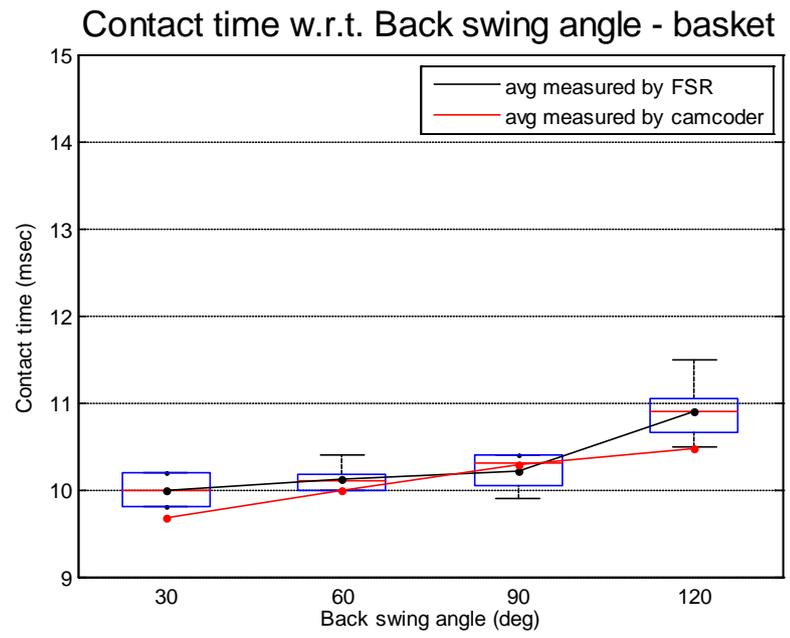

Fig. 13. Contact time with respect to back swing angle

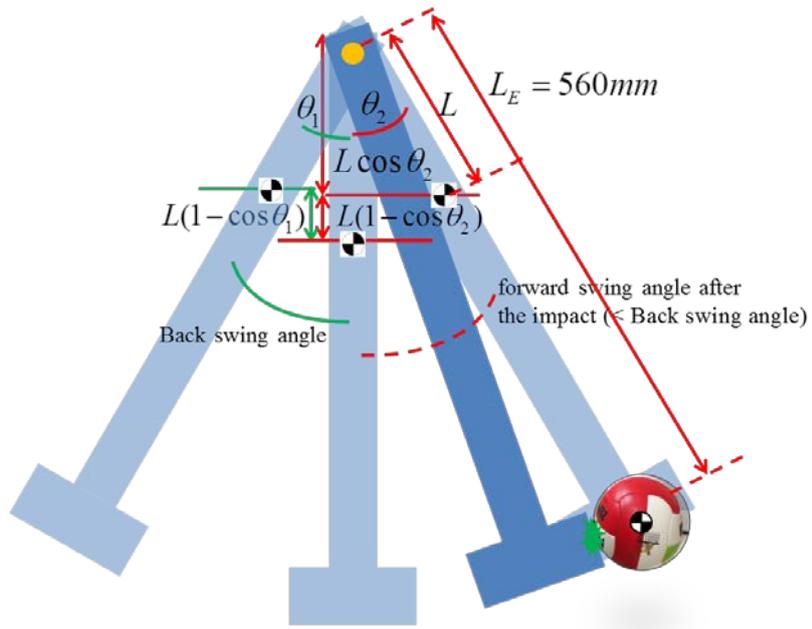

Fig. 14. Concept of solving the energy loss

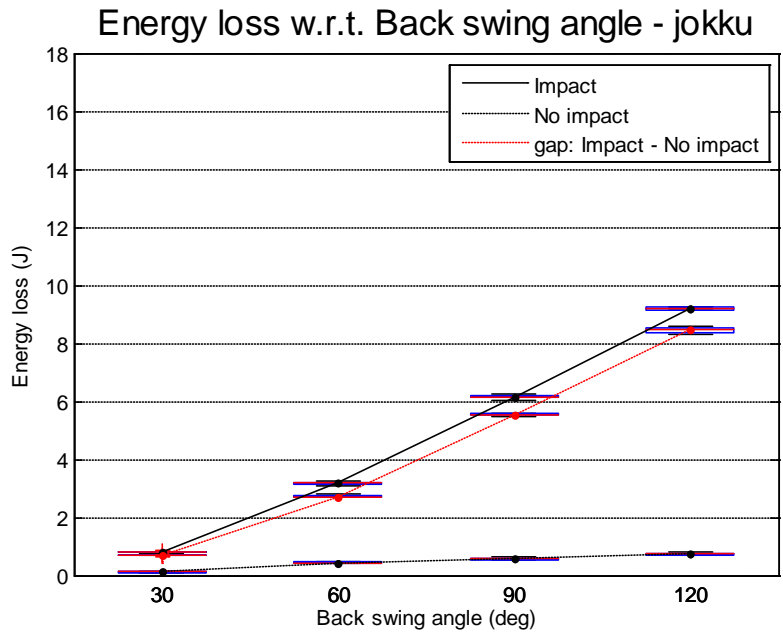

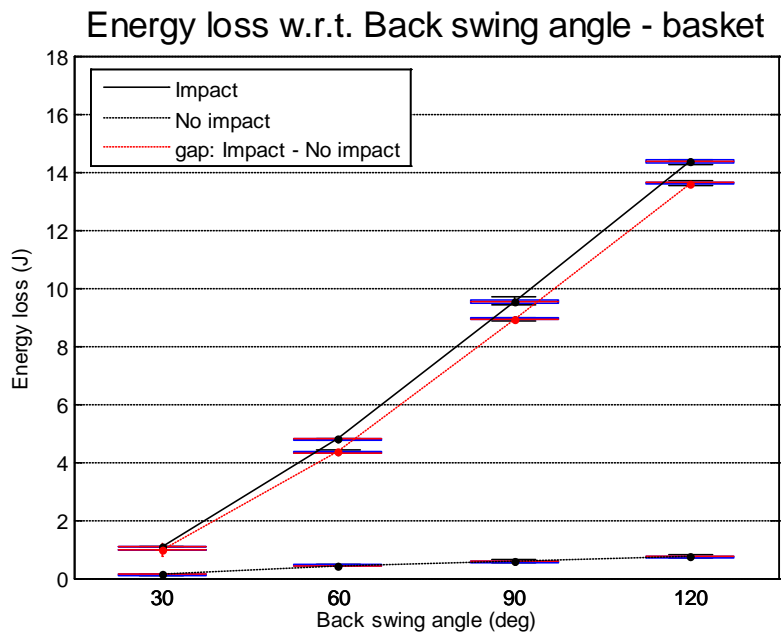

Fig. 15. Comparison of the energy loss

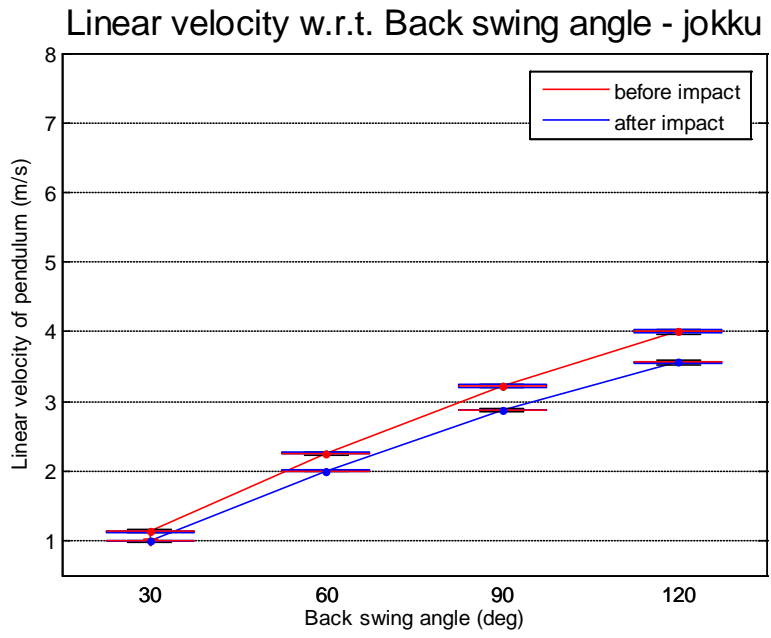

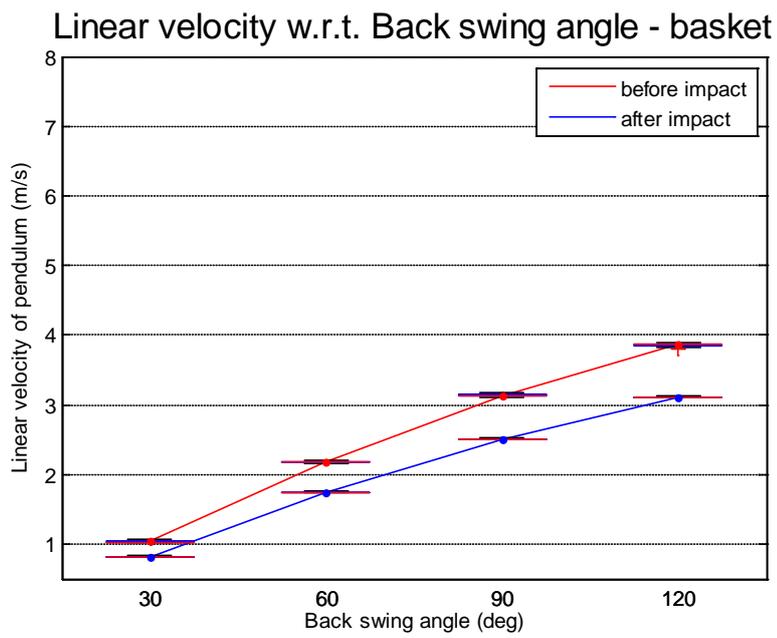

Fig. 16. Linear velocity of pendulum with respect to back swing angle

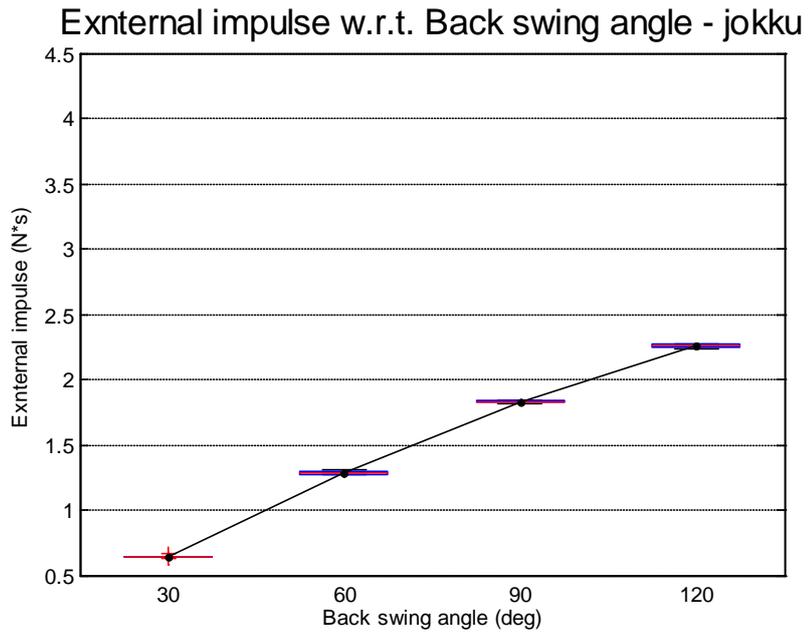

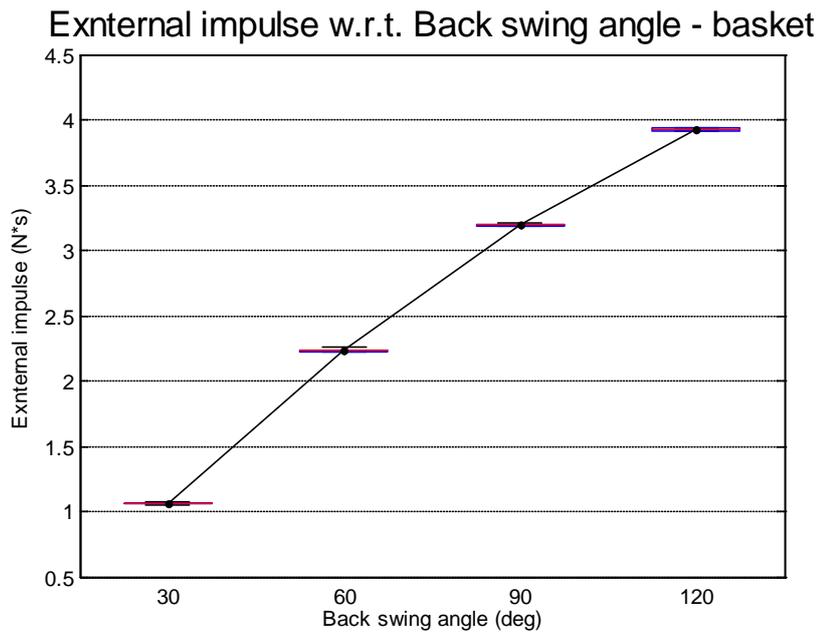

Fig. 17. External impulse with respect to back swing angle

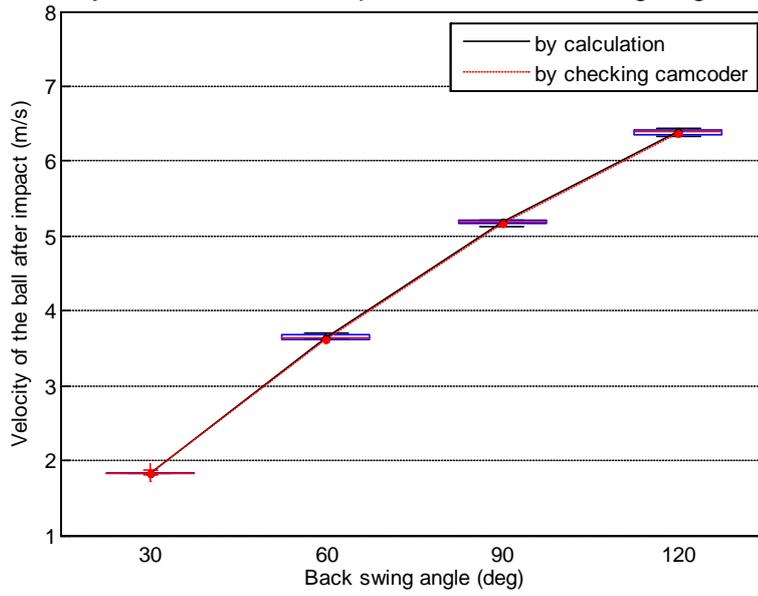

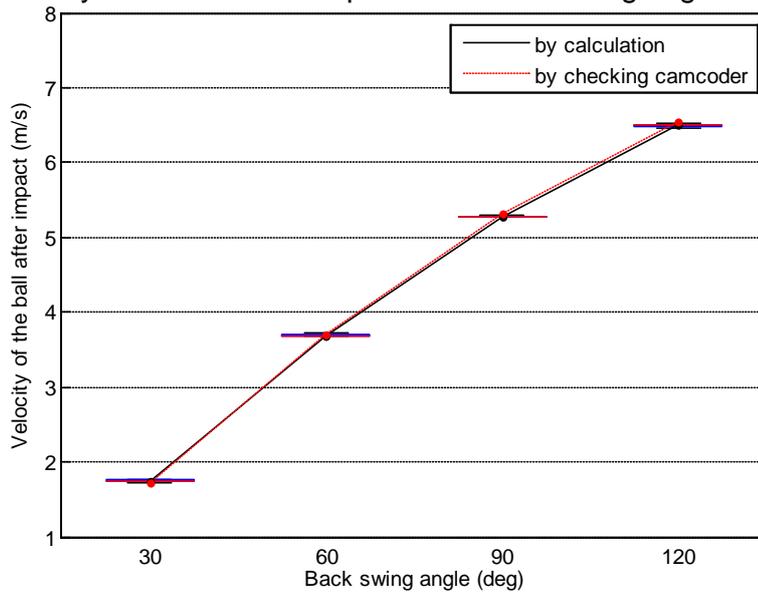

Fig. 18. Velocity of the ball after impact with respect to back swing angle

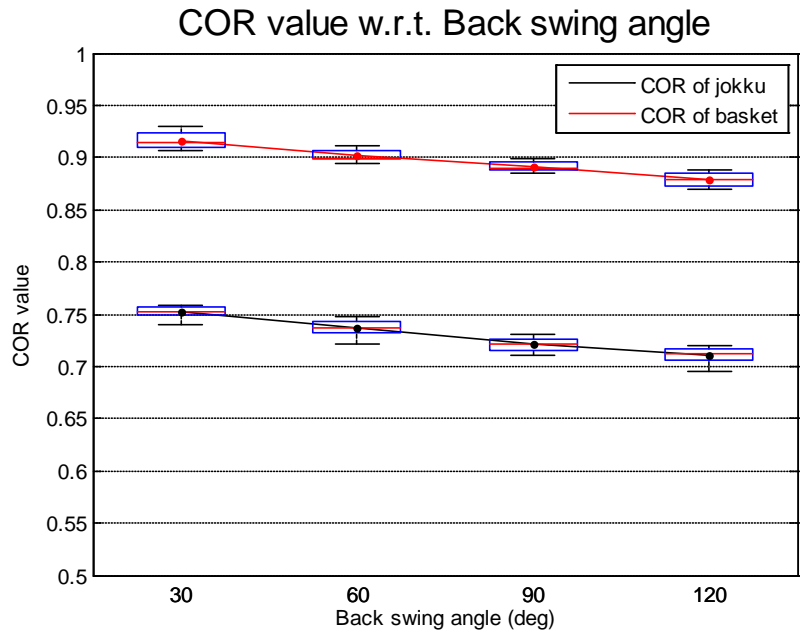

Fig. 19. COR value with respect to back swing angle

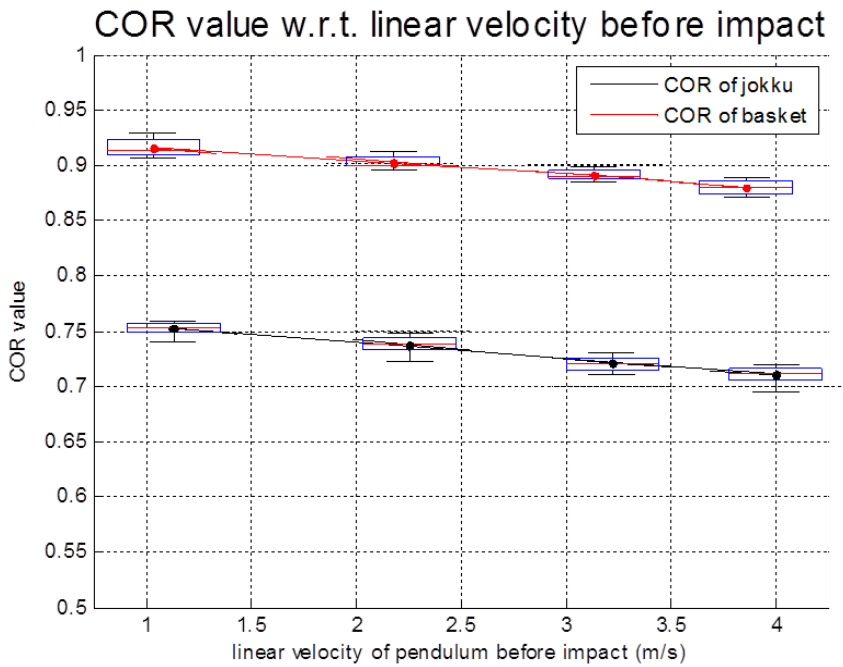

Fig. 20. COR value with respect to linear velocity of pendulum before collision

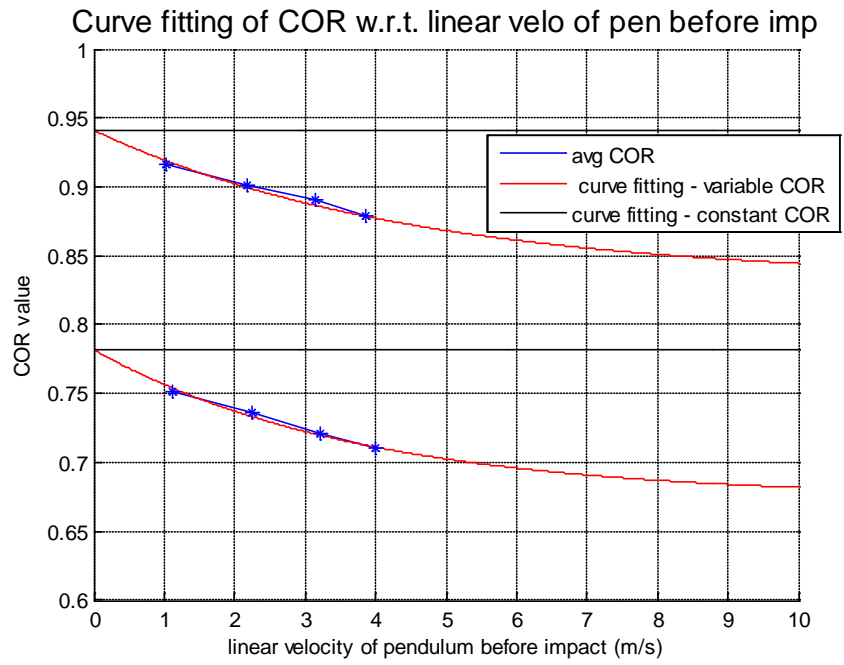

Fig. 21. Curve fitting of COR with respect to linear velocity of pendulum before collision

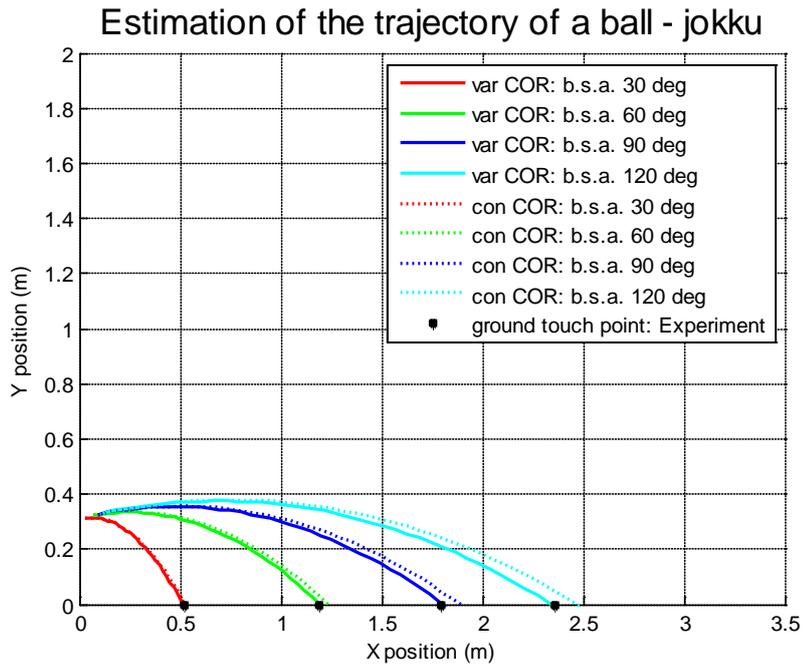

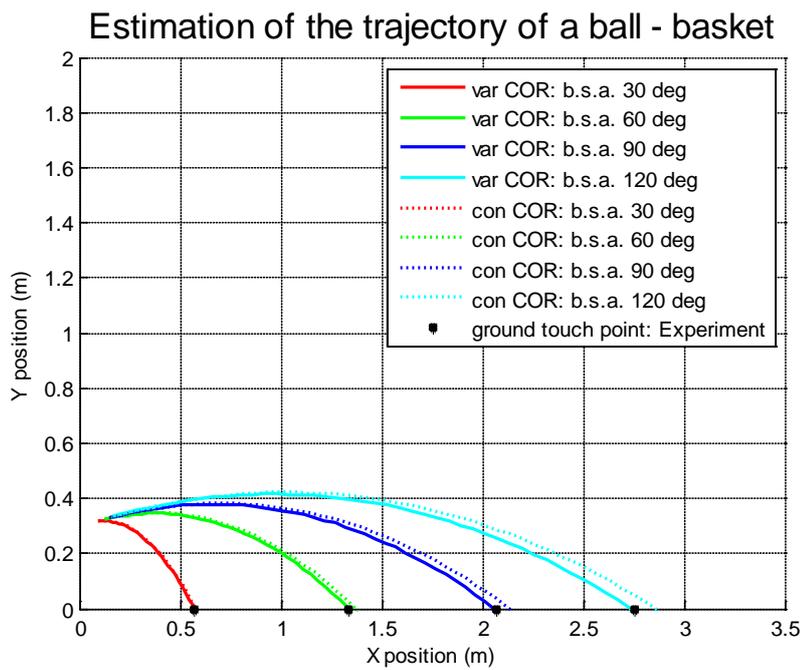

Fig. 22. Simulation of the ball travel tendency for variable and constant COR

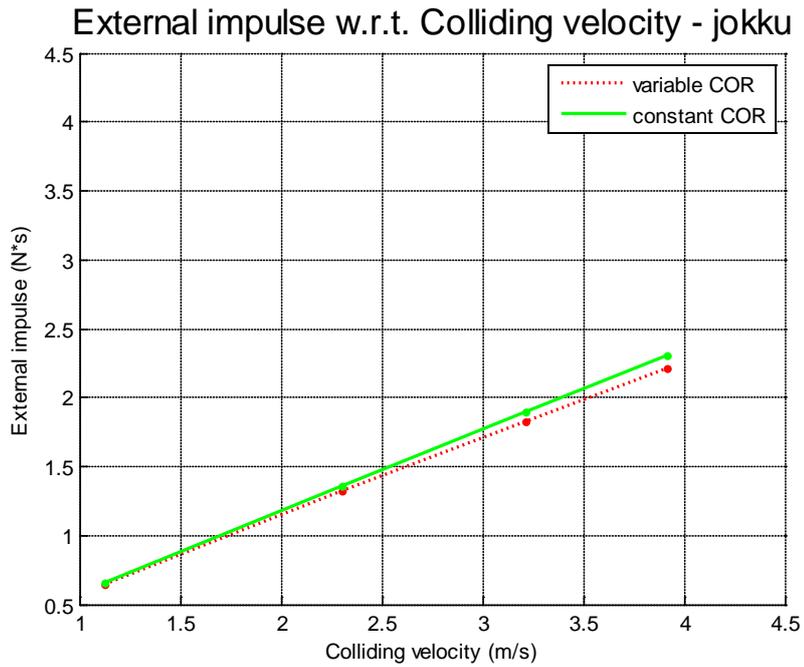

Fig. 23. Comparison of external impulse with respect to the collision velocity– jokku ball

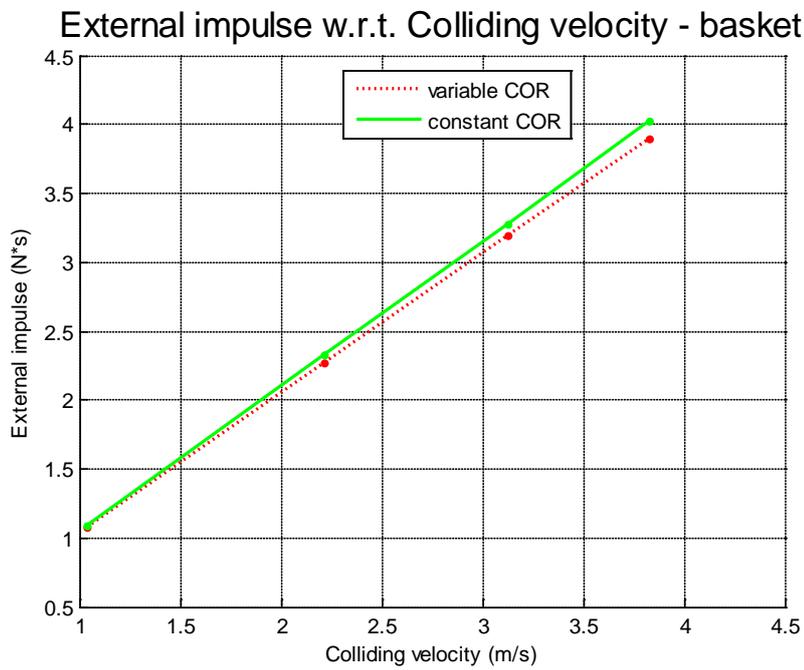

Fig. 24. Comparison of external impulse with respect to the collision velocity– basket ball

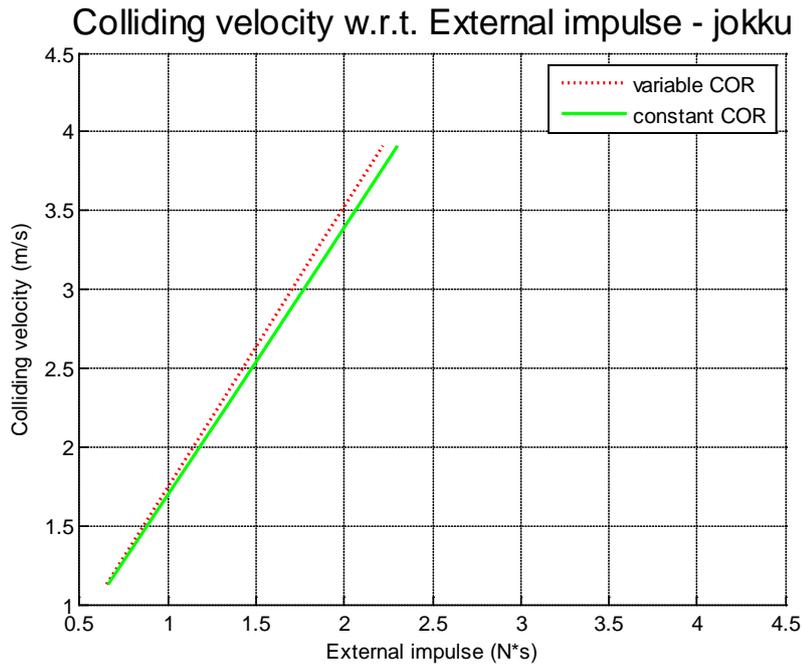

Fig. 25. Trend of comparison of collision velocity with respect to the external impulse – jokku ball

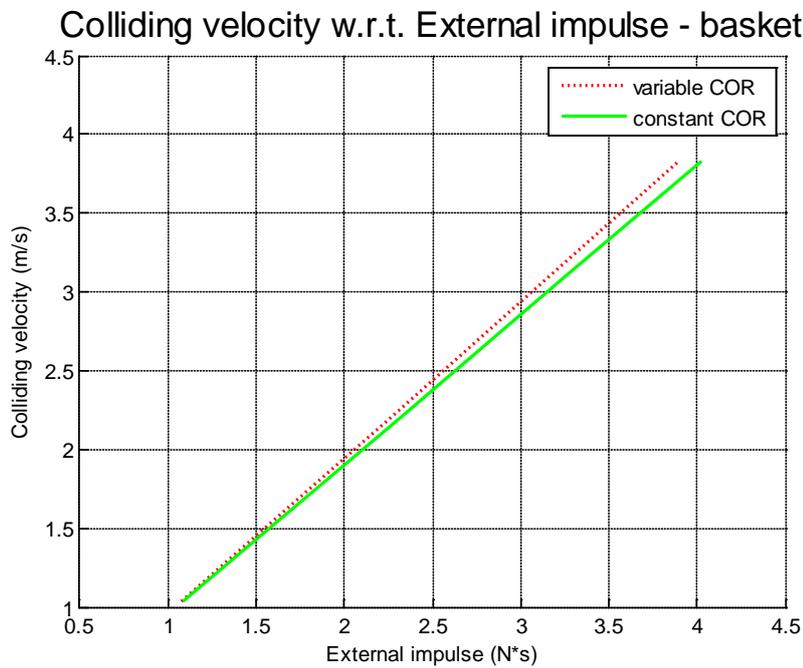

Fig. 26. Trend of comparison of collision velocity with respect to the external impulse – basket ball

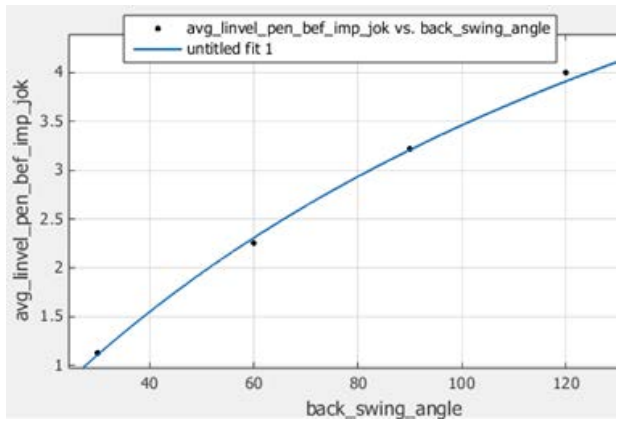 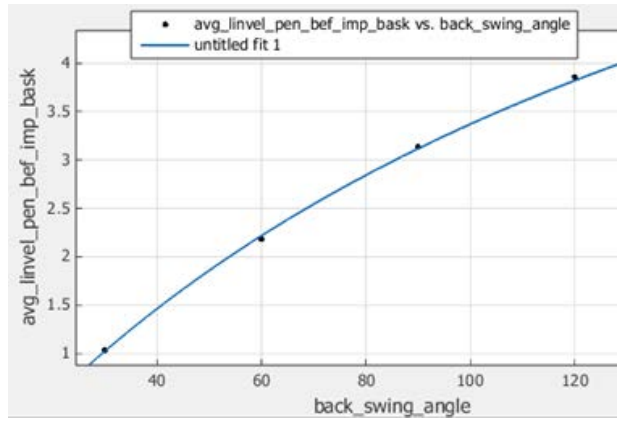

(a) Curve fitting for jokku ball  (b) Curve fitting for basket ball

Fig. 27. Trend of linear velocity of pendulum before impact (linear pen. vel. bef. imp) with respect to back swing angle (b.s.a)

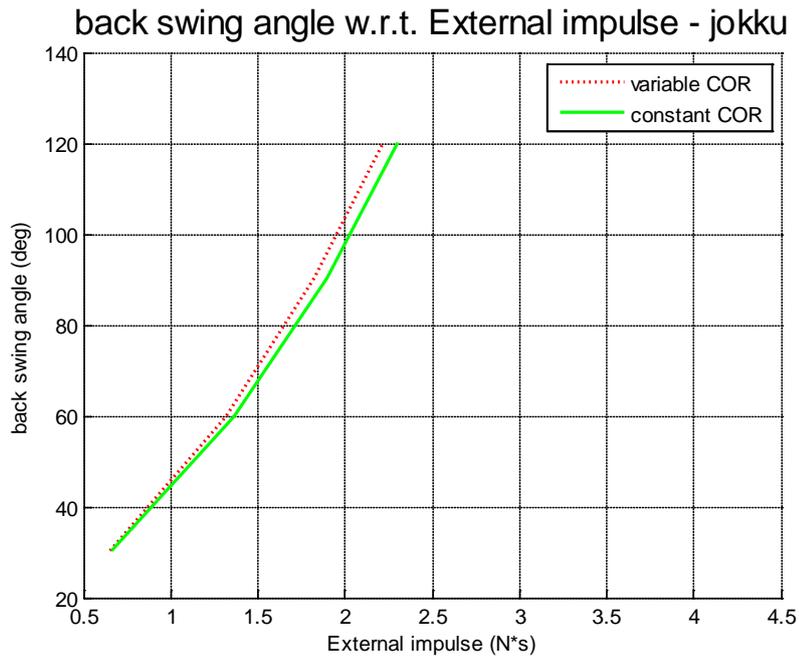

Fig. 28. Trend of comparison of back swing angle with respect to the external impulse – jokku ball

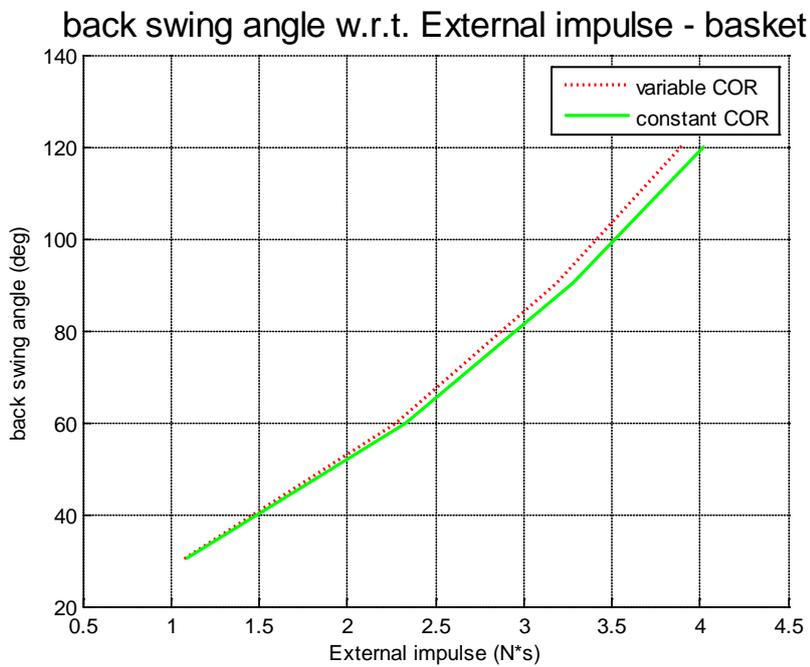

Fig. 29. Trend of comparison of back swing angle with respect to the external impulse – basket ball

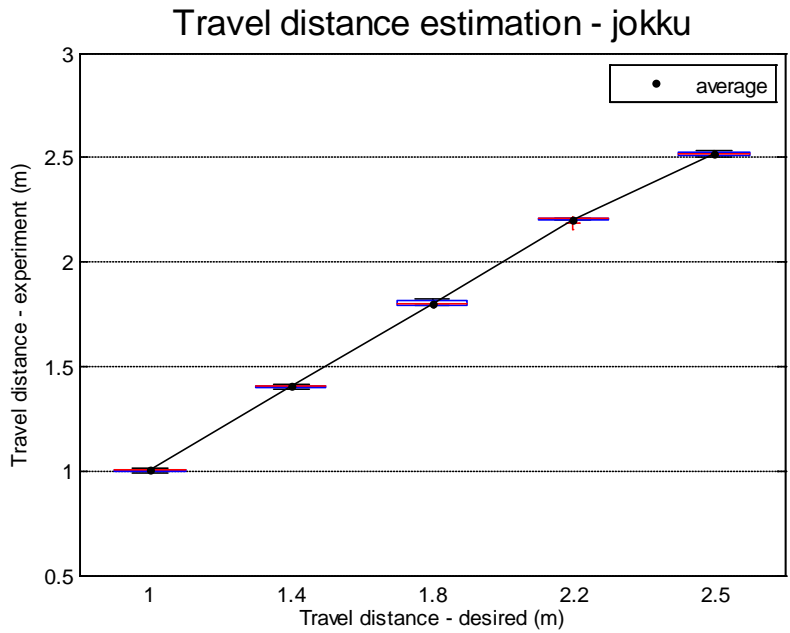

Fig. 30. Trend of comparison of travel distance between desired and experiment – jokku ball

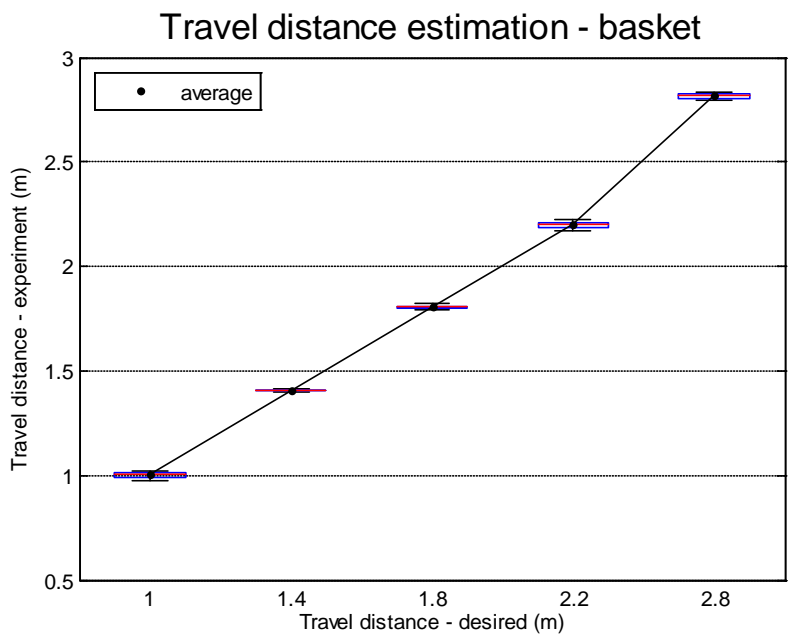

Fig. 31. Trend of comparison of travel distance between desired and experiment – basket ball